\documentclass[11pt]{article}

\usepackage[T1]{fontenc}
\usepackage{ccfonts}
\usepackage[italic]{mathastext}

\usepackage[margin=1.00in]{geometry}
\usepackage{mathtools}
\mathtoolsset{showonlyrefs}
\usepackage{amssymb}
\usepackage{xcolor}
\usepackage[skip=0.5em plus 2pt minus 1pt, indent=0pt]{parskip}

\usepackage{sectsty}
\allsectionsfont{\normalfont\scshape}

\usepackage{natbib}
\usepackage[colorlinks=true,allcolors=blue!55!black]{hyperref}

\title{\scshape Publishing Without Journals:\\ An Open, Forkable Archive with Attributed Review}
\author{Matthew Lorig\\
  \small Department of Applied Mathematics, University of Washington\\
  \small \texttt{mlorig@uw.edu}}
\date{\today}

\begin{document}
\maketitle

\begin{abstract}
\noindent
The journal is a seventeenth-century technology asked to do four modern jobs at
once: disseminate results, certify their quality, allocate scholarly attention,
and confer career credit. It does none of them well. Pre-publication peer
review is slow, only weakly reliable, demonstrably biased toward established
authors and institutions, and expensive, while the reviewing effort it consumes
is spent largely on work that will never matter. We argue that these are not
defects to be patched but consequences of bundling dissemination and
certification into a single gated act, and we propose unbundling them. Under the
proposal, authors deposit papers in an open archive; certification happens
\emph{after} deposit, continuously, through attributed and up- or down-voted
public commentary to which authors may reply; and papers are version-controlled
objects that any qualified reader may \emph{fork}, so that the lineage of an
idea---and hence the credit for it---is recorded automatically. None of the
individual components is speculative: each already exists somewhere in the
scholarly ecosystem. The contribution here is to argue that assembling them
into a single venue that \emph{replaces} rather than supplements the journal is
both feasible and preferable, and to confront the objections---sparse
participation, the chilling effect of real names, the loss of the certification
signal, and the non-meritocratic distribution of attention---that any honest
version of the argument must answer.
\end{abstract}

\section{Introduction}

Almost every complaint a working researcher has about publishing traces back to
a single design decision made when journals were invented: dissemination and
certification are welded together. A result becomes public and becomes
``peer-reviewed'' in the same act, gated by a handful of referees and one
editor. That bundling made sense when printing and postage were the binding
constraints and when a page in a bound volume was the only way to reach a
distant colleague. It makes far less sense now, when dissemination is
essentially free and instantaneous, and when the certification the bundle
provides is slow, noisy, and biased in measurable ways.

This paper argues that the two functions should be separated, and sketches a
system in which they are. Authors deposit papers in a public archive of the
arXiv or SSRN type. Certification is then produced continuously and in the open:
approved members of the community comment under their real names, comments and
papers can be voted up or down, and authors reply. Papers are stored as their
source (for example, \LaTeX), so that any approved reader may \emph{fork} a
paper---download its source, extend or correct it, and publish the derivative
with an automatic, permanent link back to its parent. The journal, as a gate, is
removed. What remains of its useful functions---quality signal, attention
allocation, credit---is reconstructed from the archive's own activity.

The rest of this paper proceeds as follows. Section~\ref{sec:failures} documents 
four ways the current system fails. Section~\ref{sec:proposal}
proposes and alternative system. Section~\ref{sec:case} sets out the benefits of the proposed system: faster consensus, less gatekeeping bias, lower cost, better-targeted scrutiny, and automatic credit for ideas.. Section~\ref{sec:priorart} shows that the design is an assembly of parts
that already work in isolation, which is the strongest available evidence of
feasibility. Section~\ref{sec:objections} addresses potential criticisms of the proposal.

\section{Four failures of the journal system}
\label{sec:failures}

\subsection{Delay}

The interval between finishing a result and having it certified is measured in
months to years, and much of that interval is pure overhead: an editor searching
for referees, referees postponing, revision rounds serialized one after another,
and finally a production queue \citep{bjork2013,huisman2017}. Surveys of authors
put typical turnaround at several months even in fast fields, with first-decision
times that have grown over the decades rather than shrunk despite the move to
electronic submission \citep{huisman2017}. In a survey of authors in
conservation biology, the median reported turnaround was fourteen weeks against a
perceived reasonable time of six, with individual horror stories running past a
year \citep{nguyen2015}; the broader picture across fields is similar, and
mathematics, computer science, and economics sit at the slow end
\citep{huisman2017,powell2016}. The cost of this delay is not merely
inconvenience. Priority disputes, stalled dissertations, and grant and job
timelines all hinge on the certification date, and knowledge that could be built
on sits idle while it waits in a queue.

\subsection{Unreliability and bias}

If review were slow but accurate, the delay might be a price worth paying. It is
not especially accurate. The classic demonstration remains that of
\citet{peters1982}, who resubmitted twelve already-published articles to the same
prestigious psychology journals that had printed them, with the authors'
prestigious affiliations replaced by fictitious low-status ones. Only three of
thirty-eight editors and reviewers noticed the papers had appeared before; of the
nine that then went through review, eight were rejected, most on grounds of
``serious methodological flaws'' in work the same journals had recently judged
excellent. Two decades later, \citet{rothwell2000} found that agreement among
reviewers in clinical neuroscience was scarcely greater than chance. The pattern
recurs wherever it is measured directly: the 2014 and 2021 consistency
experiments at the NeurIPS conference, which sent a subset of submissions through
two independent committees, found that roughly half of accepted papers would have
been rejected had the process simply been rerun \citep{beygelzimer2023}. A
process whose accept/reject verdict is that sensitive to the luck of the referee
draw is closer to a noisy lottery than to a measurement.

The noise is not even unbiased. In a controlled experiment at WSDM~2017,
single-blind reviewers---those who could see author identities---bid for and
favored papers from famous authors and prestigious institutions significantly
more than double-blind reviewers evaluating the same papers \citep{tomkins2017}.
The wider literature on reviewer bias documents effects tied to gender,
nationality, language, and seniority \citep{lee2013}. A former editor of the
\textit{BMJ}, reviewing the evidence, concluded that peer review is slow,
expensive, largely a lottery, and poor at detecting either error or fraud
\citep{smith2006}. The point is not that referees are careless; it is that two or
three of them, chosen under time pressure and reading in private, are a
statistically thin instrument on which to hang a binary, career-defining verdict.

\subsection{Cost}

The system is also expensive, and the expense is decoupled from the service
rendered. A small number of commercial publishers control more than half of all
research output and sustain operating margins in excess of thirty percent
\citep{lariviere2015,vannoorden2013}---margins that persist precisely because the
labor that creates the value, writing and refereeing, is donated by academics,
while the resulting content is sold back to their institutions through
subscriptions or to their funders through article-processing charges
\citep{vannoorden2013}. Universities pay twice, once in salaries to produce the
work and again in fees to read it. Whatever one thinks of the profit, the
architecture is hard to defend: the marginal cost of distributing a paper is now
near zero, yet the price is not.

\subsection{Misallocated effort}

Finally, the reviewing effort the system extracts is spent in the wrong place.
Every submission is refereed with roughly equal intensity before anyone knows
whether it matters, which means enormous scrutiny is lavished on work that will
never be read or cited, while the community's collective attention is rationed by
the accident of which journal accepted a paper rather than by the paper's actual
interest. Referee time is scarce and fatigue is real \citep{nguyen2015}; spending
it uniformly, up front, and in private is close to the least efficient possible
allocation. It would be better to let attention find the work that warrants it
and to concentrate scrutiny there.

\section{The proposal}
\label{sec:proposal}

The remedy is to unbundle dissemination from certification and to make
certification a continuous, open, post-deposit process. Concretely, the system
has four components.

\begin{enumerate}
\item
\emph{Open deposit.} Authors upload papers to a public archive, as they already
do on arXiv and SSRN. Deposit is the act of publication; there is no gate and no
delay. A paper is timestamped and citable the moment it appears.
\item
\emph{Attributed, votable commentary.} Each paper carries a threaded comment
section. Approved members of the community---see Section~\ref{sec:objections} on
what ``approved'' should mean---comment under their real names. Comments, and
papers, can be voted up or down by approved users, so that the signal of what the
community finds sound, important, or flawed is produced continuously and visibly
rather than compressed into one editor's private accept/reject bit.
\item
\emph{Author response.} Authors reply to comments in the same thread. Review
becomes a public conversation with a permanent record, rather than an exchange of
confidential letters that no future reader ever sees.
\item
\emph{Forking with provenance.} Because papers are stored as source, any approved
reader may fork one: take its \LaTeX, extend, correct, or build on it, and
publish the derivative with an automatic, permanent citation link to its parent.
The archive thus records the lineage of ideas the way a version-control system
records the lineage of code. Credit for an idea attaches to the node where it
first appeared, and the graph of who-built-on-whom is machine-readable.
\end{enumerate}

The journal's four jobs are then redistributed rather than abolished.
Dissemination is handled by deposit; certification by commentary and votes;
attention allocation by the vote-weighted visibility of the archive; and credit
by the provenance graph. No single gatekeeper stands between a result and its
readers.

\section{The affirmative case}
\label{sec:case}

The advantages of the proposal described in Section \ref{sec:proposal} are as follows:

\begin{enumerate}
\item
\emph{Faster consensus.} Feedback arrives in days from the whole relevant
community rather than in months from two or three referees. Because commentary is
public and cumulative, agreement (or disagreement) about a result becomes visible
as it forms, and the scientific consensus that actually drives progress is
reached and recorded more quickly.
\item
\emph{Less gatekeeping bias.} No small panel can quietly suppress a result or
wave through a friend's. The biases documented in Section~\ref{sec:failures}
depend on privacy and on a thin panel; opening the evaluation to the community and
attaching names to it removes both the cover and the bottleneck. Errors and
merits are argued in the open, where they can be weighed by anyone.
\item
\emph{Lower cost.} Authors pay no publication fee and institutions pay no
subscription. A public archive with a commentary layer is inexpensive to run
relative to the tens of billions the current system extracts
\citep{lariviere2015,vannoorden2013}, and nothing of scientific value is lost in
the saving.
\item
\emph{Better-allocated scrutiny.} Attention concentrates on work the community
finds interesting; work with no impact is simply not commented on, and no
referee-hours are spent manufacturing a verdict nobody needed. Scrutiny follows
interest instead of preceding it uniformly.
\item
\emph{Automatic credit and less plagiarism.} The provenance graph makes the
theft of an idea conspicuous, because building on prior work is done by forking
it, which records the link. Priority is established by timestamp at deposit
rather than by the vagaries of a publication queue, and derivative work carries
its ancestry with it.
\end{enumerate}

\section{This is not utopian}
\label{sec:priorart}

The strongest evidence that the design would work is that every one of its parts
already works somewhere; what is missing is only their assembly into a single
venue meant to replace the journal. Preprint archives have already decoupled
dissemination from certification at scale: arXiv and SSRN make deposit the act of
publication, and in several fields the preprint is already the object of record.
Post-publication commentary exists: PubPeer has hosted attributed and anonymous
commentary on published work since 2012, and has been the venue for a large share
of error- and fraud-detection in the recent literature. Open, published review
exists: the machine-learning conferences conduct their reviewing on OpenReview,
where reviews and rebuttals are visible, and in 2022--2023 the journal
\textit{eLife} abolished accept/reject decisions altogether, publishing every
reviewed paper as a ``reviewed preprint'' accompanied by its public reviews and
an editorial assessment---explicitly, in its editor's words, relinquishing the
journal's gatekeeping role \citep{eisen2022}. Forking and provenance exist:
Octopus (a Jisc-backed platform) breaks research into eight linked publication
types chained by their dependencies, with no accept/reject step and
community-visible review, while ResearchEquals lets authors publish and build on
modular research steps \citep{dhar2023}. The proposal here is the union of these:
an open archive whose certification, attention, and credit functions are all
produced by attributed community activity, forking included, with no journal
left in the loop.

That prior art also sharpens the contribution. Octopus and eLife keep an
editorial layer; PubPeer bolts commentary onto a literature still gated by
journals; OpenReview serves a gated conference. The claim here is more
aggressive---that the gate can be removed entirely and its functions
reconstructed from the archive---and precisely because it is more aggressive it
owes the objections a serious hearing.

\section{Objections, and how the design answers them}
\label{sec:objections}

\subsection{Will anyone actually comment?}

This is the strongest objection, and the evidence is sobering. Open commentary is
a public good, and public goods are under-supplied: the private cost of writing a
careful review is borne by the reviewer, while the benefit is diffuse. The
current system papers over this by conscripting referees through editors; remove
the conscription and voluntary participation may collapse. The data bear this
out. A study of all 2020 bioRxiv and medRxiv preprints found that only
about $7\%$ received even a single comment over roughly seven months
\citep{carneiro2023}, consistent with earlier estimates below ten percent
\citep{malicki2021}. A system in which ninety-plus percent of papers get no
feedback has not solved the certification problem; it has merely relocated it.

The design must therefore treat participation as the central engineering
problem, not an afterthought. Three levers help. First, make commentary and
review \emph{count}: attributed reviews and highly-upvoted comments should be
citable, first-class research outputs that appear on a CV, so that the effort
buys recognition rather than only costing time \citep{leek2011}. Second, retain a
light editorial nudge---a small rotating pool that solicits a minimum number of
reviews for each deposited paper---so that the baseline is not zero; this is a
far cheaper editorial function than the current one and does not gate
publication. Third, exploit the fact that, when comments do appear, they are
substantive: the same bioRxiv analysis found that the comments left were
predominantly criticisms, corrections, and suggestions of the kind traditional
review produces \citep{carneiro2023}. The scarcity is of quantity, not quality,
which is an incentive problem, and incentive problems are tractable.

\subsection{Real names and the chilling of criticism}

The proposal requires real names, and real names cut two ways. They confer
accountability and make credit assignable, but they also deter exactly the
criticism the system needs most: a junior researcher who publicly faults a senior
one risks retaliation at the next grant panel or hiring committee. It is not an
accident that PubPeer, the most active venue for catching serious problems in the
literature, permits anonymity. A pure real-name policy may therefore purchase
civility at the price of candor.

The resolution is not to abandon real names but to be selective about where they
are required. Authorship and forking, where credit is the point, should be
attributed. Critical commentary can be offered under a persistent, verified
pseudonym: an identity that is stable and reputation-bearing (so that a track
record of sound criticism accrues, and abuse can be sanctioned) but not
trivially linkable to a career. This preserves accountability at the level of the
account while protecting the critic, and it lets the community weight a comment by
the commenter's history rather than only by their name.

\subsection{The lost certification signal}

Journal prestige, for all its faults, does real work: hiring committees, tenure
cases, and grant panels use ``published in $X$'' as a compressed, portable quality
signal, and they use it precisely because they cannot read every candidate's work
closely. An archive with no accept/reject bit removes that signal, and the reform
will not be adopted until something replaces it \citep{lariviere2015}. The design
does supply replacements---vote counts, the number and standing of substantive
reviews, fork counts and downstream citations in the provenance graph---but these
are richer and noisier than a single journal name, and evaluators would have to
learn to read them. There is also a genuine risk of trading one flawed proxy
(journal brand) for another (a popularity score). The honest position is that the
signal can be reconstructed, probably better, but that doing so requires
deliberate metric design and a cultural shift in how committees evaluate, and that
this is the reform's true bottleneck---sociological, not technical.

\subsection{Attention is not meritocratic}

The affirmative case assumes that interesting work attracts scrutiny and
unimportant work is harmlessly ignored. Attention, though, is distributed by a
rich-get-richer dynamic at least as much as by merit: famous authors, prestigious
institutions, well-networked labs, and fashionable topics draw eyes regardless of
quality, and a vote count amplifies this Matthew effect rather than correcting it.
The bias documented for referees \citep{tomkins2017} does not vanish when the
crowd replaces the panel; it may intensify, since the crowd sees author names by
default. So ``unimportant papers are simply ignored'' is too sanguine: important
work by unknown authors may also be ignored, and visible endorsement may track
prominence rather than correctness. Mitigations exist---default-blinded
presentation of author identity, vote-weighting that discounts pile-on and
rewards reviewer track record, surfacing under-examined deposits---but they must
be built in deliberately. Left to a raw vote, the system risks reproducing the
inequities it set out to cure.

\subsection{Manipulation, brigading, and popularity {\normalfont$\neq$} correctness}

Any votable public forum invites gaming: coordinated up-voting of allies,
down-voting of rivals, sock-puppetry, and the general fact that a popularity
score measures agreement, not truth---and the crowd has been confidently wrong
before. ``Approved users'' is the first line of defense: participation should
require a verified scholarly identity (for example, an institutional or ORCID-
backed account), which raises the cost of Sybil attacks and makes sanctions
meaningful. Beyond that, votes should be weighted by reviewer standing rather
than counted raw, coordinated voting should be detectable and penalized, and---
crucially---the vote should be treated as an input to human judgment, not a
verdict. The system's job is to surface argument and evidence for readers to
weigh, not to compute a truth value by majority.

\subsection{Fraud and harmful dissemination}

Removing the pre-publication gate means flawed, fraudulent, or dangerous work
reaches the public immediately, as the COVID-19 preprint episode illustrated.
This is a real cost, and it should be stated plainly rather than waved away. Two
things bound it. First, the gate the reform removes was never good at catching
fraud in the first place \citep{smith2006}; open, attributed, post-deposit
scrutiny by the whole community has a substantially better recent record at
detection than closed pre-publication review did. Second, the design can retain
targeted safeguards without a general gate: clear preprint-style labeling of
review status, expedited community flagging, and heightened checks for
categories where premature dissemination carries public-health or safety risk.
The claim is not that an open archive eliminates bad work---nothing does---but
that it exposes and corrects it faster and more visibly than the system it
replaces.

\subsection{Who approves the ``approved users''?}

Finally, the word ``approved'' hides a governance problem: whoever controls
admission and moderation controls a gate, and a reform premised on removing gates
cannot smuggle one back in unexamined. The answer is to keep the admission
criterion broad, objective, and non-discretionary---a verified scholarly identity
rather than a curator's judgment---and to make moderation policy transparent,
appealable, and ideally federated across institutions rather than vested in a
single operator. Governance is where a system like this most easily goes wrong,
and it deserves as much design attention as the commentary and forking
mechanisms themselves.

\section{A migration path}
\label{sec:migration}

None of this need arrive at once, and it will not arrive by fiat. The realistic
path is incremental and rides on infrastructure that already exists. Begin by
adding an attributed, votable commentary and forking layer atop an existing
archive, so deposit and discussion coexist with conventional journals and nothing
is lost by trying it. Let scholarly societies---whose imprimatur, unlike a
commercial publisher's, is a reputational rather than a financial asset---lend
their names to curated, vote- and review-derived ``collections,'' reconstructing
a portable quality signal without a gate. Encourage funders and departments,
which set the incentives that ultimately govern behavior, to recognize
attributed reviews and provenance-graph impact in evaluation. Each step is
useful on its own, and the endpoint---an archive that has absorbed the journal's
functions---is reached by accretion rather than by asking the community to leap.

\section{Conclusion}

The journal bundles dissemination and certification because, four centuries ago,
it had to. It no longer has to, and keeping the bundle now costs the research
community time, accuracy, money, and mis-spent effort, in each case measurably
\citep{peters1982,rothwell2000,tomkins2017,beygelzimer2023,lariviere2015,huisman2017}.
Unbundling them---open deposit for dissemination, and continuous, attributed,
forkable community activity for certification, attention, and credit---addresses
all four failures at their common root. The components are proven in isolation;
the remaining work is assembly, incentive design, and governance. The hardest
problems are not technical but social: getting scholars to review when no editor
compels them, protecting critics who speak under scrutiny, and teaching
evaluators to read a provenance graph where they once read a journal's name. Those
are hard, but they are the right problems to be working on, and they are more
tractable than continuing to defend a gate that the evidence no longer supports.

\bibliographystyle{abbrvnat}
\bibliography{refs}

@article{peters1982,
  author  = {Peters, Douglas P. and Ceci, Stephen J.},
  title   = {Peer-review practices of psychological journals: The fate of published articles, submitted again},
  journal = {Behavioral and Brain Sciences},
  year    = {1982},
  volume  = {5},
  number  = {2},
  pages   = {187--195},
  doi     = {10.1017/S0140525X00011183}
}

@article{rothwell2000,
  author  = {Rothwell, Peter M. and Martyn, Christopher N.},
  title   = {Reproducibility of peer review in clinical neuroscience: Is agreement between reviewers any greater than would be expected by chance alone?},
  journal = {Brain},
  year    = {2000},
  volume  = {123},
  number  = {9},
  pages   = {1964--1969},
  doi     = {10.1093/brain/123.9.1964}
}

@article{lee2013,
  author  = {Lee, Carole J. and Sugimoto, Cassidy R. and Zhang, Guo and Cronin, Blaise},
  title   = {Bias in peer review},
  journal = {Journal of the American Society for Information Science and Technology},
  year    = {2013},
  volume  = {64},
  number  = {1},
  pages   = {2--17},
  doi     = {10.1002/asi.22784}
}

@article{smith2006,
  author  = {Smith, Richard},
  title   = {Peer review: a flawed process at the heart of science and journals},
  journal = {Journal of the Royal Society of Medicine},
  year    = {2006},
  volume  = {99},
  number  = {4},
  pages   = {178--182},
  doi     = {10.1177/014107680609900414}
}

@article{tomkins2017,
  author  = {Tomkins, Andrew and Zhang, Min and Heavlin, William D.},
  title   = {Reviewer bias in single- versus double-blind peer review},
  journal = {Proceedings of the National Academy of Sciences},
  year    = {2017},
  volume  = {114},
  number  = {48},
  pages   = {12708--12713},
  doi     = {10.1073/pnas.1707323114}
}

@article{beygelzimer2023,
  author  = {Beygelzimer, Alina and Dauphin, Yann N. and Liang, Percy and Wortman Vaughan, Jennifer},
  title   = {Has the machine learning review process become more arbitrary as the field has grown? The {NeurIPS} 2021 consistency experiment},
  journal = {arXiv preprint arXiv:2306.03262},
  year    = {2023},
  doi     = {10.48550/arXiv.2306.03262}
}

@article{lariviere2015,
  author  = {Larivi{\`e}re, Vincent and Haustein, Stefanie and Mongeon, Philippe},
  title   = {The oligopoly of academic publishers in the digital era},
  journal = {PLOS ONE},
  year    = {2015},
  volume  = {10},
  number  = {6},
  pages   = {e0127502},
  doi     = {10.1371/journal.pone.0127502}
}

@article{vannoorden2013,
  author  = {Van Noorden, Richard},
  title   = {Open access: The true cost of science publishing},
  journal = {Nature},
  year    = {2013},
  volume  = {495},
  number  = {7442},
  pages   = {426--429},
  doi     = {10.1038/495426a}
}

@article{nguyen2015,
  author  = {Nguyen, Vivian M. and Haddaway, Neal R. and Gutowsky, Lee F. G. and Wilson, Alexander D. M. and Gallagher, Austin J. and Donaldson, Michael R. and Hammerschlag, Neil and Cooke, Steven J.},
  title   = {How long is too long in contemporary peer review? Perspectives from authors publishing in conservation biology journals},
  journal = {PLOS ONE},
  year    = {2015},
  volume  = {10},
  number  = {8},
  pages   = {e0132557},
  doi     = {10.1371/journal.pone.0132557}
}

@article{huisman2017,
  author  = {Huisman, Janine and Smits, Jeroen},
  title   = {Duration and quality of the peer review process: the author's perspective},
  journal = {Scientometrics},
  year    = {2017},
  volume  = {113},
  number  = {1},
  pages   = {633--650},
  doi     = {10.1007/s11192-017-2310-5}
}

@article{powell2016,
  author  = {Powell, Kendall},
  title   = {Does it take too long to publish research?},
  journal = {Nature},
  year    = {2016},
  volume  = {530},
  number  = {7589},
  pages   = {148--151},
  doi     = {10.1038/530148a}
}

@article{bjork2013,
  author  = {Bj{\"o}rk, Bo-Christer and Solomon, David},
  title   = {The publishing delay in scholarly peer-reviewed journals},
  journal = {Journal of Informetrics},
  year    = {2013},
  volume  = {7},
  number  = {4},
  pages   = {914--923},
  doi     = {10.1016/j.joi.2013.09.001}
}

@article{leek2011,
  author  = {Leek, Jeffrey T. and Taub, Margaret A. and Pineda, Fernando J.},
  title   = {Cooperation between referees and authors increases peer review accuracy},
  journal = {PLOS ONE},
  year    = {2011},
  volume  = {6},
  number  = {11},
  pages   = {e26895},
  doi     = {10.1371/journal.pone.0026895}
}

@article{eisen2022,
  author  = {Eisen, Michael B. and Akhmanova, Anna and Behrens, Timothy E. and Diedrichsen, J{\"o}rn and Harper, Diane M. and Iordanova, Mihaela D. and Weigel, Detlef and Zaidi, Mone},
  title   = {Scientific Publishing: Peer review without gatekeeping},
  journal = {eLife},
  year    = {2022},
  volume  = {11},
  pages   = {e83889},
  doi     = {10.7554/eLife.83889}
}

@article{dhar2023,
  author  = {Dhar, Payal},
  title   = {Octopus and {ResearchEquals} aim to break the publishing mould},
  journal = {Nature},
  year    = {2023},
  doi     = {10.1038/d41586-023-00861-0}
}

@article{carneiro2023,
  author  = {Carneiro, Clarissa F. D. and da Costa, Gabriel Gon{\c{c}}alves and Neves, Kleber and Abreu, Mariana Boechat and Tan, Pedro Batista and Ray{\^e}e, Danielle and Boos, Fl{\'a}via Zacouteguy and Andrejew, Roberta and Lubiana, Tiago and Mali{\v{c}}ki, Mario and Amaral, Olavo Bohrer},
  title   = {Characterization of comments about {bioRxiv} and {medRxiv} preprints},
  journal = {JAMA Network Open},
  year    = {2023},
  volume  = {6},
  number  = {8},
  pages   = {e2331410},
  doi     = {10.1001/jamanetworkopen.2023.31410}
}

@article{malicki2021,
  author  = {Mali{\v{c}}ki, Mario and Costello, Joseph and Alperin, Juan Pablo and Maggio, Lauren A.},
  title   = {From amazing work to {I} beg to differ: analysis of {bioRxiv} preprints that received one public comment till September 2019},
  journal = {Biochemia Medica},
  year    = {2021},
  volume  = {31},
  number  = {2},
  pages   = {020201},
  doi     = {10.11613/BM.2021.020201}
}

\end{document}